\documentclass[twocolumn,showpacs,preprintnumbers,amsmath,amssymb]{revtex4}

\usepackage{graphicx}
\usepackage{braket}

\newtheorem{thm}{Theorem}

\begin{document}

\preprint{APS/123-QED}

\title{Adaptive Entanglement Purification Protocols \\ with Two-way Classical Communication}

\author{Alan W. Leung}
 \email{leung@math.mit.edu}
\author{Peter W. Shor}%
 \email{shor@math.mit.edu}
\affiliation{%
Department of Mathematics, Massachusetts Institute of Technology,
Cambridge, MA 02139, USA
}%

\date{\today}

\begin{abstract}
We present a family of entanglement purification protocols that
generalize four previous methods, namely the recurrence method,
the modified recurrence method, and the two methods proposed by
Maneva-Smolin and Leung-Shor. We will show that this family of
protocols have improved yields over a wide range of initial
fidelities F, and hence imply new lower bounds on the quantum
capacity assisted by two-way classical communication of the
quantum depolarizing channel. In particular, the yields of these
protocols are higher than the yield of universal hashing for F
less than 0.993 and as F goes to 1.
\end{abstract}

\maketitle

\section{Introduction}

Quantum information theory studies the information processing
power one can achieve by harnessing quantum mechanical principles.
Many important results such as quantum teleportation, superdense
coding, factoring and search algorithms make use of quantum
entanglements as fundamental resources\cite{BBCJPW, BW, S, G}.
Pure-state entanglements are therefore useful; however, when they
are exposed to noise, they become mixed entangled states. It is
thus important to study the procedures by which we can extract
pure-state entanglements from mixed entangled states, and we call
these procedures entanglement purification protocols(EPP). The
present work in particular studies the scenario where the two
parties - whom we call Alice and Bob throughout - are allowed to
communicate classically. We will follow the framework of
\cite{BDSW, MS, LS1} and generalize these results to obtain a
family of protocols with improved yields.

\section{Adaptive Entanglement Purification Protocols (AEPP)}

\subsection{Notations}

We denote von Neumann entropy by $S(\rho)$, Shannon entropy by
$H(p_0,p_1,\ldots)$ and label the four Bell states with two
classical bits $(a,b)$ as follows:

\begin{eqnarray}
00:&\ket{\Phi^{+}}=\frac{1}{\sqrt{2}}(\ket{\uparrow\uparrow}+\ket{\downarrow\downarrow})\nonumber\\
01:&\ket{\Psi^{+}}=\frac{1}{\sqrt{2}}(\ket{\uparrow\downarrow}+\ket{\downarrow\uparrow})\nonumber\\
10:&\ket{\Phi^{-}}=\frac{1}{\sqrt{2}}(\ket{\uparrow\uparrow}-\ket{\downarrow\downarrow})\nonumber\\
11:&\ket{\Psi^{-}}=\frac{1}{\sqrt{2}}(\ket{\uparrow\downarrow}-\ket{\downarrow\uparrow}).\nonumber
\end{eqnarray}

\noindent This work concerns the purification of the generalized
Werner state\cite{W},

\begin{eqnarray}
\rho_F=&&F \ket{\Phi^{+}}\bra{\Phi^{+}}+\frac{1-F}{3}\bigg(
\ket{\Phi^{-}}\bra{\Phi^{-}}\nonumber\\
&& + \ket{\Psi^{+}}\bra{\Psi^{+}} +
\ket{\Psi^{-}}\bra{\Psi^{-}}\bigg).\nonumber
\end{eqnarray}

\noindent At the beginning of these entanglement purification
protocols, two persons, Alice and Bob, share a large number of
quantum states $\rho_F$, say $\rho_F^{\otimes N}$, and they are
allowed to communicate classically, apply unitary transformations
and perform projective measurements. We place no restriction on
the size of their ancilla systems so that we lost no generality in
restricting their local operations to unitaries and projective
measurements. In the end, the quantum states $\Upsilon$ shared by
Alice and Bob are to be a close approximation of the maximally
entangled states $(\ket{\Phi^{+}} \bra{\Phi^{+}})^{\otimes M}$, or
more precisely we require the fidelity between $\Upsilon$ and
$(\ket{\Phi^{+}} \bra{\Phi^{+}})^{\otimes M}$ approaches zero as
$N$ goes to infinity. We define the yield of such protocols to be
$M/N$.

We will often use the BXOR operation by which we mean the
bilateral application of the two-bit quantum XOR (or
controlled-NOT). We only consider the scenario in which Alice and
Bob share two (or more) bipartite quantum states that are Bell
diagonal and they apply BXOR to two pairs of quantum states such
that one pair is the ``source'' and one pair is the ``target''.
Using the two classical bit notations, we write

\begin{eqnarray}
BXOR(i,j): \{0,1\}^N &\rightarrow & \{0,1\}^N \nonumber\\
(a_i,b_i) &\mapsto & (a_i\oplus a_j, b_i) \nonumber\\
(a_j,b_j) &\mapsto & (a_j,b_i \oplus b_j) \nonumber\\
(a_k,b_k) &\mapsto & (a_k,b_k) \textrm{ if $k\neq i, j$}\nonumber
\end{eqnarray}

\noindent when Alice and Bob share N pairs of bipartite quantum
states and they apply BXOR to the ith pair as source and the jth
pair as target.

\subsection{\label{sec-2B}Description of AEPP}
\begin{figure}[h]
\includegraphics[width=40mm]{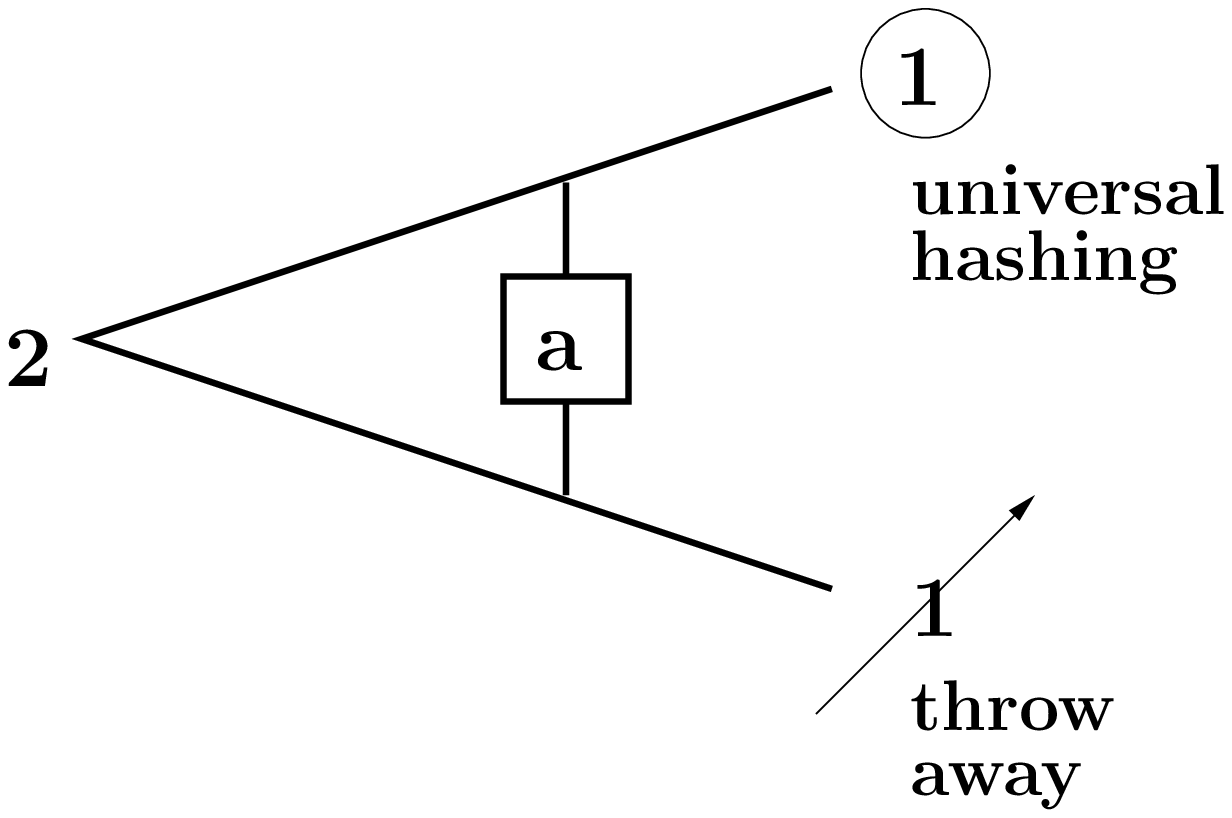}
\caption{\label{pic-1}AEPP(a,2)}
\end{figure}

\noindent 1. AEPP(a,2): Alice and Bob put the bipartite quantum
states $\rho_F^{\otimes N}$ into groups of two, apply BXOR(1,2)

$$(a_1,b_1,a_2,b_2)\mapsto(a_1\oplus a_2, b_1, a_2, b_1\oplus
b_2)$$

\noindent and take projective measurements on the second pair
along the z-axis. Using two-way classical communication channel,
they can compare their measurement results. If the measurement
results agree($b_1\oplus b_2 = 0$), then it is likely that there
has been no amplitude error and Alice and Bob will perform
universal hashing on the first pair; if the results
disagree($b_1\oplus b_2 = 1$), they throw away the first pair
because it is likely that an amplitude error has occurred. We give
a graphical representation of this protocol in fig.\ref{pic-1}.

\noindent 2. AEPP(a,4): Alice and Bob put the bipartite quantum
states $\rho_F^{\otimes N}$ into groups of four, apply BXOR(1,4),
BXOR(2,4), BXOR(3,4)

\begin{eqnarray}
(a_1,b_1,a_2,b_2,a_3,b_3,a_4,b_4)  \mapsto \nonumber \\
(a_1\oplus a_4, b_1, a_2\oplus a_4, b_2,  a_3\oplus a_4, b_3,
a_4,b_1 \oplus b_2 \oplus b_3 \oplus b_4)\nonumber
\end{eqnarray}

\noindent and take projective measurements on the fourth pair
along the z-axis. Using two-way classical communication channel,
they can compare their measurement results. If the measurement
results agree($b_1\oplus b_2\oplus b_3\oplus b_4 =0$), then it is
likely that there has been no amplitude error and Alice and Bob
will perform universal hashing on the first three pairs together.

On the other hand, if the results disagree($b_1\oplus b_2\oplus
b_3\oplus b_4 =1$), it is likely that there is one amplitude error
and Alice and Bob want to locate this amplitude error. They do so
by applying BXOR(2,1)

\begin{eqnarray}
(a_1\oplus a_4, b_1, a_2\oplus a_4, b_2,  a_3\oplus a_4, b_3,
a_4,1)\mapsto \nonumber \\
(a_1\oplus a_4, b_1\oplus b_2, a_1\oplus a_2, b_2, a_3\oplus a_4,
b_3, a_4,1)\label{LN-1}
\end{eqnarray}

\noindent and taking projective measurements on the first pair
along the z-axis. Note that the second pair($a_1\oplus a_2, b_2$)
and the third pair($a_3\oplus a_4, b_3$) are no longer entangled.
Alice and Bob then use classical communication channel to compare
their results. If the results agree($b_1 \oplus b_2 =0$), then the
amplitude error detected by the first measurements is more likely
to be on either the third or the fourth pair than on the first
two. Therefore Alice and Bob perform universal hashing on the
second pair and throw away the third pair. If the results
disagree($b_1 \oplus b_2 =1$), then the amplitude error is more
likely to be on the first two pairs. In this case, Alice and Bob
perform universal hashing on the third pair and throw away the
second pair.

\begin{figure}[h]
\includegraphics[width=40mm]{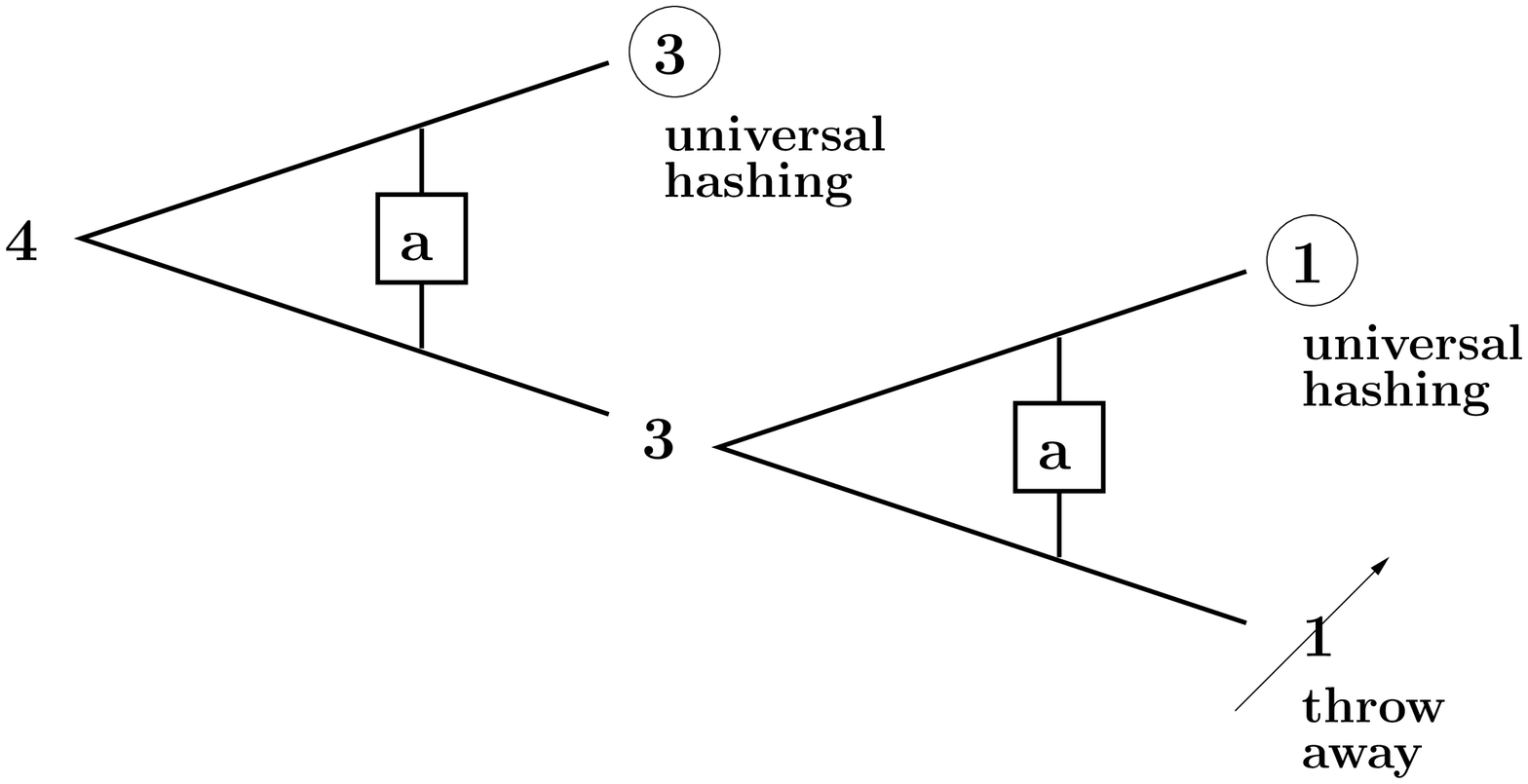}
\caption{\label{pic-2}AEPP(a,4)}
\end{figure}

Note that the amplitude error could have been on the fourth pair
but this protocol works well even if that is the case; and also
that with this procedure we always end up with one pair on which
Alice and Bob can perform universal hashing when the first
measurement results disagree. We represent this protocol
graphically in fig.\ref{pic-2}.

\noindent 3. AEPP(a,8): Alice and Bob put the bipartite quantum
states $\rho_F^{\otimes N}$ into groups of eight, apply BXOR(1,8),
BXOR(2,8), BXOR(3,8), \ldots, BXOR(7,8)

\begin{eqnarray}
(a_1,b_1,a_2,b_2,\ldots,a_7,b_7,a_8,b_8)  \mapsto \nonumber \\
(a_1\oplus a_8, b_1, a_2\oplus a_8, b_2, \ldots, a_7\oplus a_8,
b_7, a_8,b_1 \oplus  \ldots \oplus b_8)\nonumber
\end{eqnarray}

\noindent and take projective measurements on the eighth pair
along the z-axis. Using classical communication channel, Alice and
Bob compare their measurement results. If the results
agree($b_1\oplus \ldots \oplus b_8 =0$), then an amplitude error
is not likely and they perform universal hashing on the first
seven pairs together.

\begin{figure}[h]
\includegraphics[width=70mm]{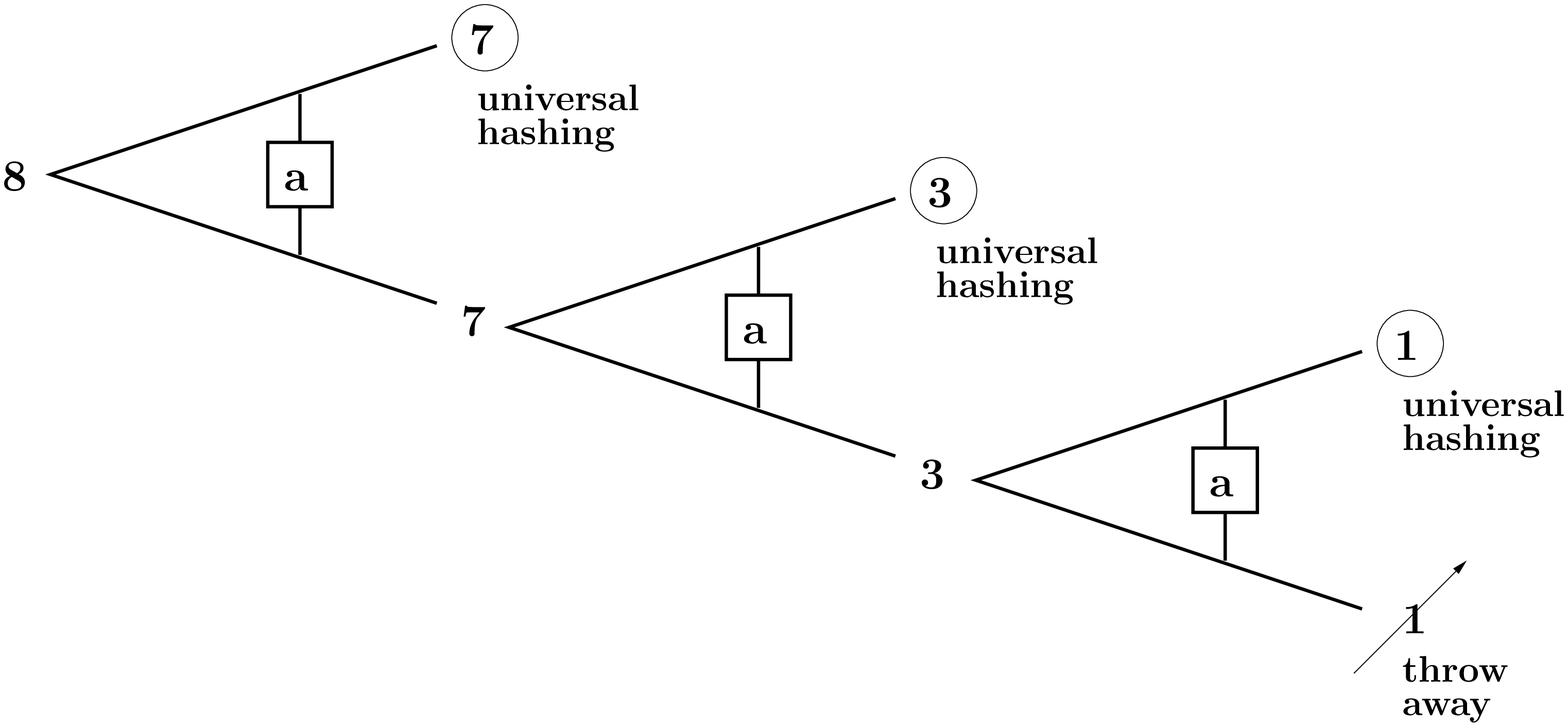}
\caption{\label{pic-3}AEPP(a,8)}
\end{figure}

On the other hand, if the measurement results disagree($b_1\oplus
\ldots \oplus b_8 =1$), then Alice and Bob want to catch this
amplitude error and they do that by applying BXOR(2,1), BXOR(3,1),
BXOR(4,1)

\begin{eqnarray}
&&(a_1\oplus a_8, b_1, a_2\oplus a_8, b_2, \ldots, a_7\oplus a_8,
b_7, a_8,1)\mapsto \nonumber\\
&&(a_1\oplus a_8, b_1 \oplus b_2\oplus b_3\oplus b_4, a_1\oplus
a_2,b_2, a_1\oplus a_3, b_3,  \nonumber\\
&&a_1\oplus a_4, b_4,a_5\oplus a_8, b_5, a_6\oplus a_8, b_6,
a_7\oplus a_8,b_7,a_8,1)\nonumber
\end{eqnarray}

\noindent and taking projective measurements on the first pair
along the z-axis. Note that the second, third and fourth pairs are
not entangled with the fifth, sixth and seventh pairs. After Alice
and Bob compare their results with classical communication channel
and if the results disagree ($b_1\oplus b_2\oplus b_3 \oplus
b_4=1$), they perform universal hashing on the fifth, sixth and
seventh pairs because $b_1\oplus b_2\oplus b_3 \oplus b_4=1$ and
$b_1\oplus \ldots \oplus b_8 =1$ together imply $b_5\oplus b_6
\oplus b_7 \oplus b_8 =0$. The first four pairs are now
represented by $(a_1\oplus a_8, 1, a_1\oplus a_2, b_2, a_1\oplus
a_3, b_3, a_1\oplus a_4, b_4)$, and it can be easily seen that we
are in the same situation as the left hand side of equation
(\ref{LN-1}): Alice and Bob know that $b_1\oplus b_2\oplus b_3
\oplus b_4 =1$ and the pair on which they measured to find out
this information has its phase error added to the other three
pairs. Therefore Alice and Bob can apply the same procedure as
equation (\ref{LN-1}) and end up with one pair that they will
perform universal hashing on. Now if the results actually agree
($b_1\oplus b_2\oplus b_3 \oplus b_4=0$), the same procedure still
applies but we need to switch the roles played by the first four
pairs and by the last four pair. We represent this protocol
graphically in fig.\ref{pic-3}.

\begin{figure}[h]
\includegraphics[width=85mm]{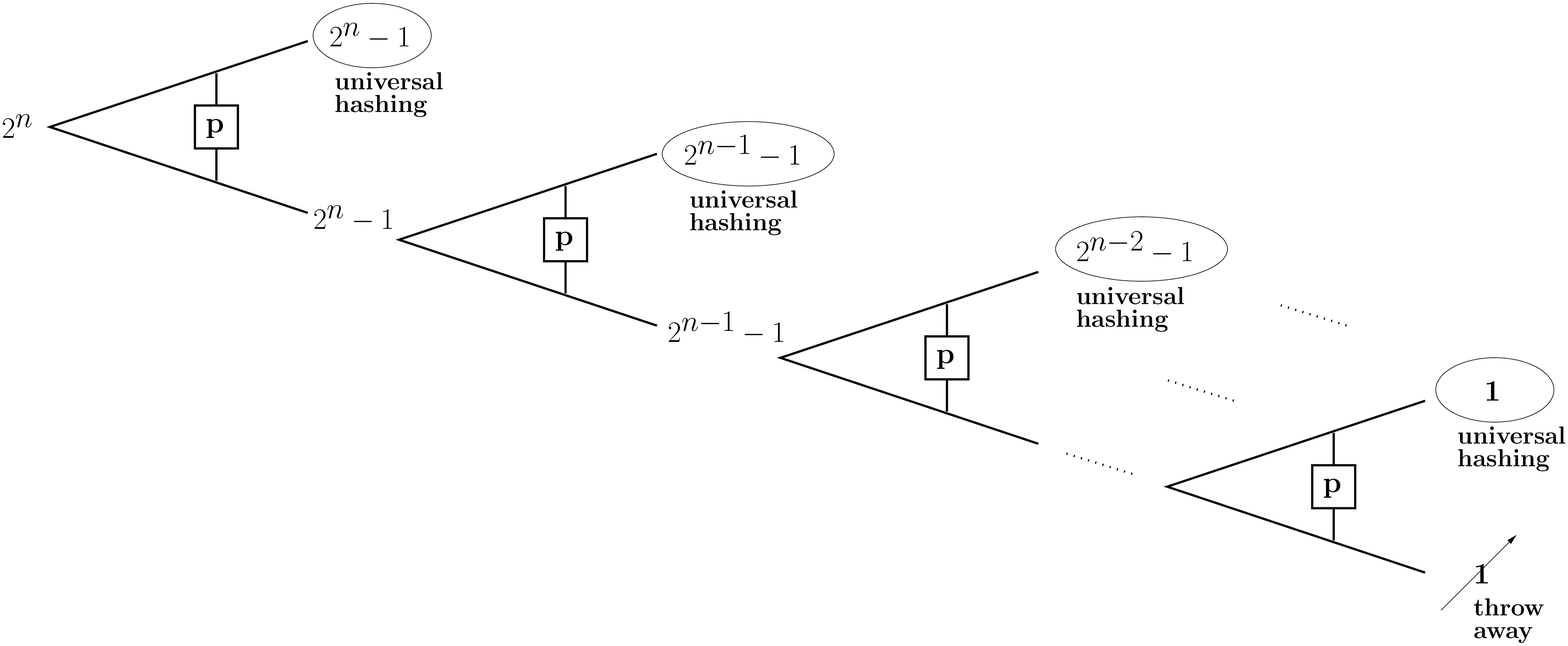}
\caption{\label{pic-4}AEPP(p,N=$2^n$)}
\end{figure}

\begin{figure}
\includegraphics[width=80mm]{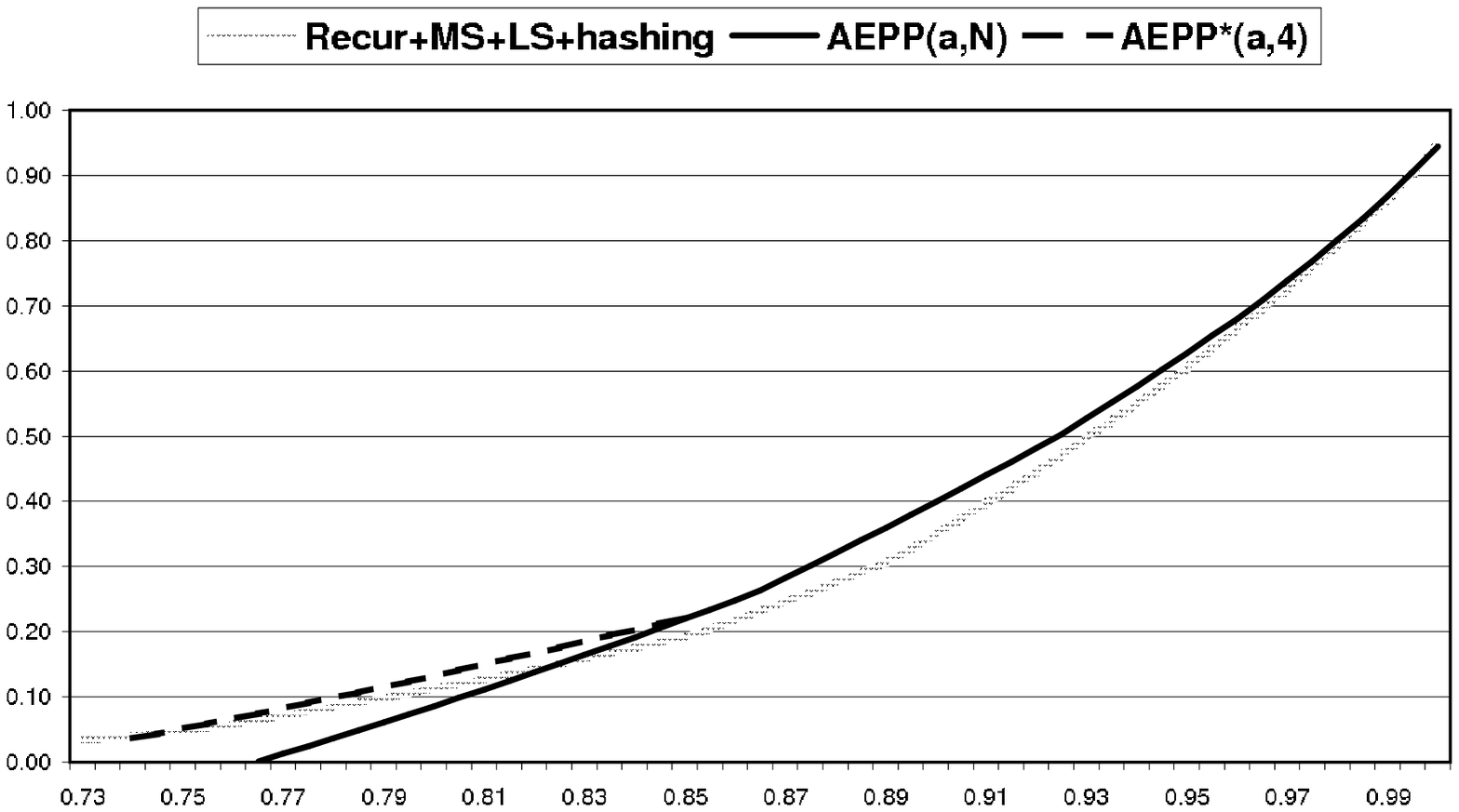}
\caption{\label{pic-5}Comparison of AEPP and previous methods: The
lightly colored line is the yield of the four methods discussed in
sec.\ref{sec-2B}; the solid line represents the yields of
AEPP(a,$N=2^n$) where $n=2,3,4,5,6$; the dashed line represents
the optimized AEPP(a,4), which is denoted by AEPP*(a,4)(see
sec.\ref{sec-2C} and \cite{LS2} for details)}
\end{figure}

\noindent 4. AEPP(a,N=$2^n$) and AEPP(p,N=$2^n$): Clearly, the
above procedures generalize to AEPP(a,N=$2^n$) and can be proved
inductively. The procedures - AEPP(a,N=$2^n$) - we discussed so
far focus on amplitude error. If we instead try to detect phase
error by switching the source pairs and target pairs in all the
BXOR operations and measuring along the x-axis rather than the
z-axis, AEPP(p,N=$2^n$) can be defined analogously. We represent
the protocols AEPP(p,N=$2^n$) graphically in fig.\ref{pic-4}, and
we present the yields of AEPP(a,$N=2^n$) for $n=2,3,4,5,6$ in
fig.\ref{pic-5}.

\subsection{\label{sec-2C}Generalization of previous methods}

In this section, we show that four previous protocols - the
recurrence method, the modified recurrence method and the two
methods proposed by Maneva-Smolin and Leung-Shor - all belong to
the family AEPP(a/p,$N=2^n$).

\noindent 1. The Recurrence Method: The recurrence
method\cite{BDSW} is the repeat applications of AEPP(a,2). When
Alice and Bob have identical measurement results, rather than
applying universal hashing right away, they repeatedly apply
AEPP(a,2) until it is more beneficial to switch to hashing.

\noindent 2. The modified recurrence method: The modified
recurrence method\cite{BDSW} is the repeat, alternate applications
of AEPP(a,2) and AEPP(p,2). After Alice and Bob apply AEPP(a,2)
and obtain identical measurement results, rather than applying
universal hashing right away, they repeatedly and alternately
apply AEPP(p,2), AEPP(a,2) and so forth until it becomes more
beneficial to switch to universal hashing.

\begin{figure}[h]
\includegraphics[width=35mm]{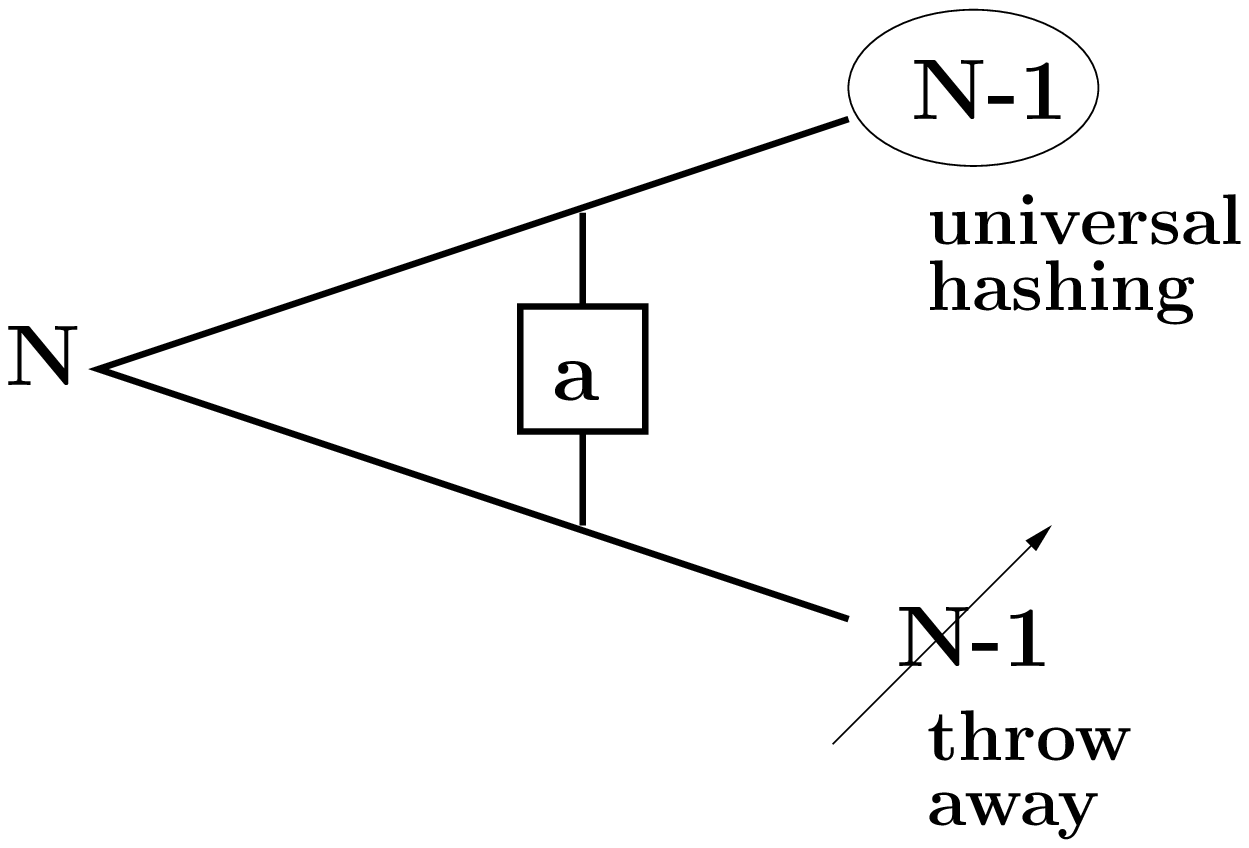}
\caption{\label{pic-6}The Maneva-Smolin method\cite{MS}}
\end{figure}

\noindent 3. The Maneva-Smolin method: The Maneva-Smolin
method\cite{MS} is to apply the first step of AEPP(a,N). Perform
universal hashing on the N-1 pairs if the measurement results
agree but throw away all the N-1 pairs if they do not. This is
illustrated in fig.\ref{pic-6}.

\begin{figure}[h]
\includegraphics[width=60mm]{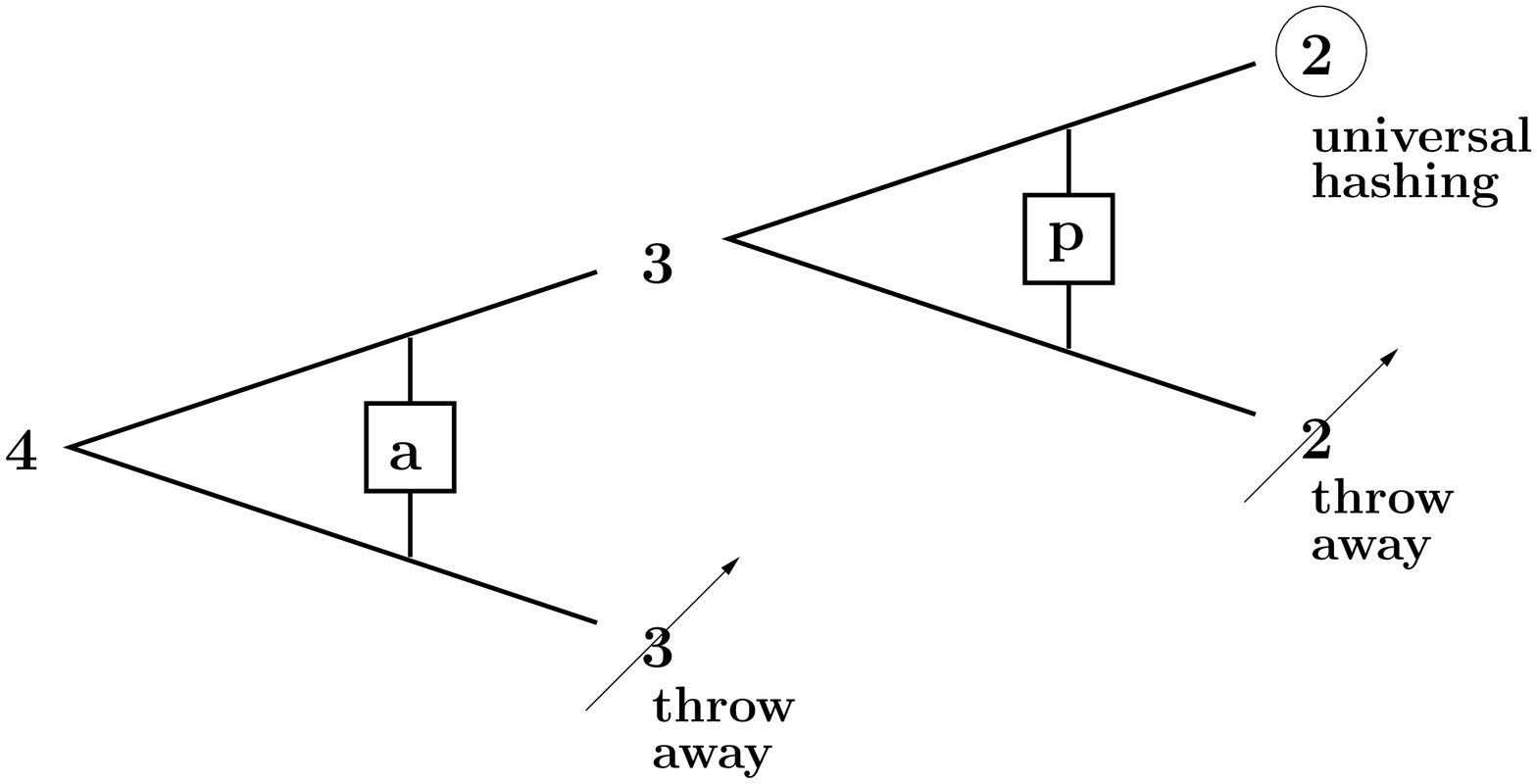}
\caption{\label{pic-7}The Leung-Shor method\cite{LS1}}
\end{figure}

\noindent 4. The Leung-Shor method: The Leung-Shor
method\cite{LS1} is a combination of the first step AEPP(a,4) and
AEPP(p,4); however, this method fails to utilize all entanglements
by throwing away the 3 pairs if the first measurement results
disagree. This is illustrated in fig.\ref{pic-7}.

\subsection{Optimization}
After we apply AEPP(a,N=$2^n$), we might end up with either
$2^n-1$ pairs or n-1 groups of pairs ($2^{n-1}-1$, $2^{n-2}-1$,
\ldots $2^k-1$, \ldots $3$ and $1$) pairs depending on the results
of the first measurements. It is possible to treat these n-1
groups differently because they are not entangled to each other.
We can either perform universal hashing(as in the Maneva-Smolin
method) or apply AEPP(p,$2^k-1$)(as in the Leung-Shor method). If
we do apply AEPP(p,$2^k-1$), we will end up with two groups of
quantum states of different sizes because we started with
$N=2^k-1$ rather than $N=2^k$. We perform optimization for
AEPP(a,4) and achieve improved yields for $F>0.74$ \cite{LS2}.

\subsection{Higher yield than universal hashing}
As we can see from fig. \ref{pic-5}, the yields of AEPP(a,$N=2^n$)
exceed the yield of universal hashing for $F<0.993$. In
\cite{LS2}, we showed the following theorem:

\begin{thm}
Let $N=2^n$ where n is a positive integer. Denote by $Y_{AEPP}$
the yields of AEPP(a,N) on the Werner state $\rho_F$. Then

{\setlength\arraycolsep{2pt}
\begin{eqnarray}
Y_{AEPP}&=&1-\frac{p}{N}(1+S_{N-1}) \nonumber \\
&&-\frac{1-p}{N}(n+1+S_{\frac{N}{2}-1}+S_{\frac{N}{4}-1}+\ldots+S_3+S_1)\nonumber
\end{eqnarray}}

\noindent where $p= \textrm{prob}(b_1 \oplus b_2 \oplus \ldots b_N
=0)$ and $S_{K-1}=H(a_1\oplus a_K, b_1, a_2\oplus a_K, b_2,
\ldots, a_{K-1}\oplus a_K,b_{K-1}|b_1\oplus \ldots \oplus b_K=0)$
for $K=2,4,8,\ldots,2^n$. Furthermore, let $F=\frac{2^n-1}{2^n}$
and $G=\frac{1-F}{3}$. Then

{\setlength\arraycolsep{2pt}
\begin{eqnarray}
\lim_{n \to \infty} Y_{AEPP}&\geq&1-H(F,G,G,G)+\frac{H(p^*)-p^*}{N} \nonumber \\
&>& 1 - H(F,G,G,G) \nonumber \\
&=& \textrm{Yield of universal hashing on $\rho_F$}\nonumber
\end{eqnarray}}

\noindent where

{\setlength\arraycolsep{2pt}
\begin{eqnarray}
p^*=\lim_{n\to \infty}p =
\frac{1}{2}(1+e^{-\frac{4}{3}}).\nonumber
\end{eqnarray}}

\end{thm}

\section{Conclusion}
We presented a family of entanglement purification protocols
AEPP(a,N) with improved yields over previous two-way entanglement
purification protocols. Moreover, the yields of these protocols
are higher than the yield of universal hashing for $F<0.993$
(shown numerically) and as F goes to 1 (shown analytically in
\cite{LS2}).

After the completion of this work, it came to our attention
similar works have been carried out in \cite{VV,HDM}.

\end{document}